\documentclass[aip,pop,reprint,superscriptaddress]{revtex4-1}

\usepackage[final]{graphicx}
\usepackage{amsmath}
\usepackage{verbatim}
\usepackage{graphicx} %include figures
\usepackage{bm} %bold math
\usepackage{dcolumn}
\usepackage{natbib}
\usepackage{subfigure}
\usepackage{sidecap}
\usepackage{soul,xcolor}
\setstcolor{red}
\usepackage[colorlinks,
linkcolor=blue,
anchorcolor=blue,
citecolor=blue,
urlcolor=blue]
{hyperref}

\begin{document}

\preprint{}

\title{Effect of thermal ions on fluid nonlinear frequency shift of ion acoustic waves in multi-ion species plasmas}

%\author{}
%\email[]{chengzhuo@pku.edu.cn}
%\thanks{}
\author{Q. S. Feng}
\affiliation{Institute of Applied Physics and Computational
	Mathematics, Beijing, 100094,  China}

\author{Q. Wang}
%\email{wangqingph@foxmail.com}
\affiliation{HEDPS, Center for
	Applied Physics and Technology, Peking University, Beijing, 100871, China}

\author{L. H. Cao}
\email{cao\_lihua@iapcm.ac.cn}
\affiliation{Institute of Applied Physics and Computational
	Mathematics, Beijing, 100094, China}
\affiliation{HEDPS, Center for
	Applied Physics and Technology, Peking University, Beijing, 100871,  China}
\affiliation{Collaborative Innovation Center of IFSA (CICIFSA), Shanghai Jiao Tong University, Shanghai 200240, China}

\author{C. Y. Zheng} \email{zheng\_chunyang@iapcm.ac.cn}
\affiliation{Institute of Applied Physics and Computational
	Mathematics, Beijing, 100094, China}
\affiliation{HEDPS, Center for
	Applied Physics and Technology, Peking University, Beijing, 100871,  China}
\affiliation{Collaborative Innovation Center of IFSA (CICIFSA), Shanghai Jiao Tong University, Shanghai 200240, China}

\author{Z. J. Liu}
\affiliation{Institute of Applied Physics and Computational
	Mathematics, Beijing, 100094, China}
\affiliation{HEDPS, Center for
	Applied Physics and Technology, Peking University, Beijing, 100871,  China}

\author{X. T. He}
\affiliation{Institute of Applied Physics and Computational
	Mathematics, Beijing, 100094, China}
\affiliation{HEDPS, Center for
	Applied Physics and Technology, Peking University, Beijing, 100871,  China}
\affiliation{Collaborative Innovation Center of IFSA (CICIFSA), Shanghai Jiao Tong University, Shanghai 200240, China}

\date{\today}

\begin{abstract}
  A model of the fluid nonlinear frequency shift of ion acoustic waves (IAWs) in multi-ion species plasmas is presented, which considers the effect of ion temperature. Because the thermal ion exists in plasmas in inertial confinement fusion (ICF) and also solar wind, which should be considered in nonlinear frequency shift of IAWs. However, the existing models [Berger et al., Physics of Plasmas 20, 032107 (2013); Q. S. Feng et al., Phys. Rev. E 94, 023205 (2016)] just consider the cold ion fluid models. This complete theory considering multi-ion species and thermal ions will calculate the frequency of the large amplitude nonlinear IAWs more accurately, especially the slow mode with high ion temperature, which will have wide application in space physics and inertial confinement fusion.

\end{abstract}

\pacs{52.35.Fp, 52.35.Mw, 52.35.Py, 52.38.Bv}
%\keywords{}

\maketitle

%{$^\S$ Q. S. Feng and Q. Wang contribute equally.}

\section{\label{Sec: Introduction}Introduction}
The nonlinearities of ion acoustic waves (IAWs) are of fundamental interest to plasma physics. Understanding the fluid effects and kinetic effects of nonlinear IAWs is of key significant in space physics such as solar wind \cite{Vecchio_2014JGR,Valentini_2014APJL,Gurnett_1977JGR,Gurnett_1978JGR,Gurnett_1979JGR} and also stimulated Brillouin scattering (SBS) in inertial confinement fusion (ICF) \cite{He_2016POP,Glenzer_2010Science,Glenzer_2007Nature,LanKe_2017PRE,Lan_2016MRE,Huo_2016PRL,Huo_2016MRE}.

The nonlinear saturation of SBS \cite{Froula_2002PRL,Berger_1998POP,Neumayer_2008PRL,Giacone_1998POP,Vu_2001PRL,Albright_2016POP} in plasmas relevant to ICF is closely related to the ion-acoustic wave saturation. Therefore, studying the nonlinearities of IAWs is important to understand the underlying physics of the saturation of SBS and to interpret scattered light levels from current ICF experiments. The nonlinear frequency shift (NFS) of IAWs induced by trapping \cite{Froula_2002PRL,Giacone_1998POP,Vu_2001PRL,Albright_2016POP} and harmonic generation \cite{Bruce_1997POP, Rozmus_1992POP} is suggested to be a possible saturation mechanism of SBS. Therefore, the theory to calculate the nonlinear frequency shift of IAWs is vital to ICF.

The nonlinear effects on the frequency of the nonlinear IAWs are hot topics as a result of their potential role in determining the saturation of SBS in ICF. \cite{Cohen_1997POP,Albright_2016POP}
The fluid NFS of IAWs resulting from harmonic generation is obtained by the isothermal cold ion fluid equations where ion is considered to be cold \cite{Berger_2013POP,Chapman_2013PRL,Feng_2016PRE}. However, in fusion plasmas, ion temperature might be almost comparable to electron temperature. Therefore, the effect of ion temperature on the fluid NFS should be considered.

In this paper, a multi thermal ion fluid model is proposed to calculate the fluid NFS of IAW in multi-ion species plasmas. This model considers the thermal ions and is verified to be consistent to Vlasov simulation data better, especially for the slow IAW mode with high ion temperature.

\section{Theoretical Model\label{sec2}}
From the isothermal hot ion fluid equations:
\begin{eqnarray}
&& \partial_t n_a+\partial_x(n_av_a)=0,  \label{eq1}\\
&& \partial_t v_a+v_a\partial_x v_a = -C_a^2\partial_x\phi-\gamma_av_{th,a}^2\partial_x n_a,\label{eq2} \\
&& -\partial^2_{x}\phi+4\pi en_{0}\exp[\phi] =4\pi\sum_a q_an_a,\label{eq3}
\end{eqnarray}
the fluid nonlinear frequency shift resulting from harmonic generation is derived.
Where the electrostatic potential is normalized by $T_e/e$, i.e., $\phi\to e\phi/T_e$; $C_a=\sqrt{Z_aT_e/m_a}$ and $v_{th,a}=\sqrt{T_a/m_a}$ are the sound speed and the thermal velocity of ion $a$. The thermal ion effect, i.e., the term of $-\gamma_av_{th,a}^2\partial_x n_a$ in Eq. (\ref{eq2}) is considered in this paper. 
Following Pesme {\it et al}. \cite{Pesme_2005POP}, Berger {\it et al}. \cite{Berger_2013POP} and Feng {\it et al}. \cite{Feng_2016PRE} the variables in a Fourier series are given by
\begin{align}
(\phi, n_a, v_a)=&(\phi_0,n_{a0},v_{a0})+\frac{1}{2}\sum_{l\neq0}(\phi_l,n_{al},v_{al})\nonumber\\
&\times\exp[il(kx-\omega t)],
\end{align}
where $(\phi_{a-l},n_{a-l},v_{a-l})=(\phi_{al},n_{al},n_{al})^*$ and $\exp\phi\simeq1+\phi+\frac{1}{2}\phi^2$ by keeping terms for $l=0,\pm 1,\pm2$ up to $2^{\mathrm{nd}}$ order. 
From Eqs. (\ref{eq1})-(\ref{eq3}) for $l=0$ by retaining terms with matching exponents in the Fourier series, one obtains
\begin{equation}
\phi_0+\frac{\phi_0^2}{2}=-\frac{\phi^2_1}{4}-\frac{\phi^2_2}{4},
\end{equation}
where conservation of charge, $n_{0}=\sum_{a}Z_an_{a0}$, has been used. 
The equation of Eq. (\ref{eq2}) for $l=1$ is
\begin{align}
%&-i\omega n_{a1}+ikn_{a0}v_{a1}=-i\frac{1}{2}n_{a-1}v_{a2}-i\frac{1}{2}kn_{a2}v_{a-1},\label{eq6}\\
&-i\omega v_{a1}+ikC_a^2\phi_1+ik\frac{\gamma_av_{th,a}^2}{n_{a0}}n_{a1}=-i\frac{1}{2}v_{a-1}v_{a2},\label{eq7}
%&[\frac{T_e}{e}k^2+4\pi n_{e0}e(1+\phi_0)]\phi_1-4\pi\sum_{a}q_an_{a1}=-\frac{1}{2}4\pi n_{e0}e\phi_2\phi_{-1},
\end{align}
where the left hand are the linear terms and the right hand are the nonlinear terms. The corresponding equation for $l=2$ is
\begin{align}
%&-2i\omega n_{a2}+2ikn_{a0}v_{a2}+ikn_{a1}v_{a1}=0,\\
&-2i\omega v_{a2}=-2ik\left[C_a^2\phi_2+\frac{1}{4}v_{a1}^2+\frac{\gamma_av_{th,a}^2}{n_{a0}}n_{a2}\right].
%&\left[\frac{4T_e}{e}k^2+4\pi n_{e0}e(1+\phi_0)\right]\phi_2+\frac{1}{4}4\pi n_{e0}e\phi_{1}^2=4\pi\sum_{a}q_{a}n_{a2}.
\label{eq4}
\end{align}
Keeping terms only to second order in $\phi_2$, we will calculate for $\phi_2$ in the following. Firstly,
\begin{align}
& n_{a2}=\frac{k}{\omega}\left[n_{a0}v_{a2}+\frac{1}{2}n_{a1}v_{a1}\right],\\
&v_{a2}=\frac{k}{\omega}\left[C_a^2\phi_2+\frac{1}{4}v_{a1}^2+\gamma_av_{th,a}^2\frac{n_{a2}}{n_{a0}}\right],
\end{align}
which further gives
\begin{equation}
\begin{aligned}
v_{a2}&=\frac{k}{\omega}\left[C_a^2\phi_2+\frac{1}{4}v_{a1}^2+\gamma_av_{th,a}^2\frac{k}{\omega}[v_{a2}+\frac{1}{2}\frac{n_{a1}}{n_{a0}}v_{a1}]\right],\nonumber\\
&=\frac{k/\omega}{1-\frac{k^2\gamma_av_{th,a}^2}{\omega^2}}[C_a^2\phi_2+\frac{1}{4}v_{a1}^2+\frac{1}{2}\frac{k\gamma_av_{th,a}^2}{\omega}\frac{n_{a1}}{n_{a0}}v_{a1}],\\
n_{a2}&=n_{a0}\frac{k^2}{\omega^2}\frac{1}{1-\frac{k^2\gamma_av_{th,a}^2}{\omega^2}}\left[C_a^2\phi_2+\frac{1}{4}v_{a1}^2+\frac{1}{2}\frac{\omega}{k}\frac{n_{a1}}{n_{a0}}v_{a1}\right].
\end{aligned}
\end{equation}

Substituting above results into  Eq. (\ref{eq4}), one obtains
\begin{align}
&[4k^2\lambda_{De}^2+(1+\phi_0)]\phi_2+\frac{1}{4}\phi_1^2\nonumber\\
=&\sum_{a}Z_a\frac{n_{a0}}{n_{e0}}\frac{k^2/\omega^2}{1-\frac{k^2\gamma_av_{th,a}^2}{\omega^2}}\left[C_a^2\phi_2+\frac{1}{4}v_{a1}^2+\frac{1}{2}\frac{\omega}{k}\frac{n_{a1}}{n_{a0}}v_{a1}
\right].\label{eq5}
\end{align}
Since $\phi_2$ is estimated to second order in $\phi_2$, $n_{a1}$ and $v_{a1}$ in Eq. (\ref{eq5}) are simply estimated to first order in $\phi_1$. Thus, from Eq. (\ref{eq7}), one obtains
\begin{align}
n_{a1}&=\frac{k}{\omega}n_{a0}v_{a1}=n_{a0}\frac{k^2/\omega^2}{1-\frac{k^2\gamma_av_{th,a}^2}{\omega^2}}C_a^2\phi_1,\\
v_{a1}&=\frac{k}{\omega}C_a^2\phi_1+\frac{k}{\omega}\gamma_av_{th,a}^2\frac{n_{a1}}{n_{a0}}=\frac{k/\omega}{1-\frac{k^2\gamma_av_{th,a}^2}{\omega^2}}C_a^2\phi_1.
\end{align}
Substituting these results into Eq. (\ref{eq5}), one obtains
\begin{align}
&[4k^2\lambda_{De}^2+(1+\phi_0)]\phi_2+\frac{1}{4}\phi_1^2\nonumber\\
=&\sum_{a}Z_a\frac{n_{a0}}{n_{e0}}\frac{k^2/\omega^2}{1-\frac{k^2\gamma_av_{th,a}^2}{\omega^2}}[C_a^2\phi_2+\frac{3}{4}v_{a1}^2].\nonumber\\
\end{align}
This result can be rewritten to
\begin{align}
&[4k^2\lambda_{De}^2+(1+\phi_0)-\sum_a\frac{Z_an_{a0}/n_{e0}}{1-\frac{k^2\gamma_av_{th,a}^2}{\omega^2}}\frac{k^2C_a^2}{\omega^2}]\phi_2\nonumber\\
=&\frac{3}{4}\sum_a\frac{Z_an_{a0}}{n_{e0}}\left[\frac{k^4C_a^4/\omega^4}{(1-\frac{k^2\gamma_av_{th,a}^2}{\omega^2})^3}-\frac{1}{3}\right]\phi_1^2.
\end{align}
Due to $\phi_0\sim|\phi_1|^2$, this term in the left hand of the above equation is neglected and 
\begin{align}
C_{2s}^2\equiv\sum_{a}\frac{Z_an_{a0}/n_{e0}}{1-\frac{k^2\gamma_av_{th,a}^2}{\omega^2}}C_a^2,\quad C_{4s}^4\equiv \sum_{a}\frac{Z_an_{a0}/n_{e0}}{(1-\frac{k^2\gamma_av_{th,a}^2}{\omega^2})^3}C_a^4,
\end{align}
is defined. Where $\frac{k^2\gamma_av_{th,a}^2}{\omega^2}$ is the term considering the thermal ion effect, which is zero in cold ion model \cite{Berger_2013POP,Feng_2016PRE}.
One obtains
\begin{align}
[4k^2\lambda_{De}^2+1-\frac{k^2C_{2s}^2}{\omega^2}]\phi_2=[\frac{3}{4}\frac{k^4C_{4s}^4}{\omega^4}-\frac{1}{4}]\phi_1^2.
\end{align}
Finally, the relation between $\phi_2$ and $\phi_1$ is
\begin{align}
\phi_2=A_{2\phi}\phi_1^2,\quad A_{2\phi}=\frac{[\frac{3}{4}\frac{k^4C_{4s}^4}{\omega^4}-\frac{1}{4}]}{[4k^2\lambda_{De}^2+1-\frac{k^2C_{2s}^2}{\omega^2}]}.
\end{align}
Defining $C_{s2}^2=4k^2C_{2s}^2/(1+4k^2\lambda_{De}^2)$, then one obtains
\begin{equation}
A_{2\phi}=\frac{C_{s2}^2}{4\omega^2-C_{s2}^2}\left[\frac{3k^2C_{4s}^4}{4\omega^2C_{2s}^2}-\frac{\omega^2}{4k^2C_{2s}^2}\right].
\end{equation}
In the following, $\phi_1$ will be estimated to less than third order in $\phi_1$ from equations for $l=0$. First, the density and velocity are
\begin{align}
n_{a1}&=\frac{k}{\omega}[n_{a0}v_{a1}+\frac{1}{2}n_{a-1}v_{a2}+\frac{1}{2}n_{a2}v_{a-1}],\\
v_{a1}&=\frac{k}{\omega}[C_a^2\phi_1+\frac{\gamma_av_{th,a}^2}{n_{a0}}n_{a1}+\frac{1}{2}v_{a-1}v_{a2}].
\end{align}
The first order and the second harmonic of density and velocity are 
\begin{align}
n_{a1}&=\frac{k}{\omega}n_{a0}v_{a1}=\frac{k^2/\omega^2}{1-\frac{k^2\gamma_av_{th,a}^2}{\omega^2}}n_{a0}C_a^2\phi_1,\\
v_{a1}&=\frac{k/\omega}{1-\frac{k^2\gamma_av_{th,a}^2}{\omega^2}}C_a^2\phi_1,\\
n_{a2}&=n_{a0}\frac{k^2/\omega^2}{1-\frac{k^2\gamma_av_{th,a}^2}{\omega^2}}
\left[C_a^2\phi_2+\frac{3}{4}v_{a1}^2\right],\\
v_{a2}&=\frac{k/\omega}{1-\frac{k^2\gamma_av_{th,a}^2}{\omega^2}}\left[C_a^2\phi_2+\frac{1}{4}v_{a1}^2+\frac{1}{2}\frac{k^2\gamma_av_{th,a}^2}{\omega^2}v_{a1}^2\right].
\end{align}
Now, we first derive velocity $v_{a1}$ by substituting result of first order into the term of high order
\begin{align}
v_{a1}
%=&\frac{k}{\omega}\left[C_a^2\phi_1+\frac{\gamma_av_{th,a}^2}{n_{a0}}\frac{k}{\omega}[n_{a0}v_{a1}+\frac{1}{2}n_{a-1}v_{a2}
%\right.\nonumber\\&\left.+\frac{1}{2}n_{a2}v_{a-1}]+\frac{1}{2}v_{a-1}v_{a2}\right],\nonumber\\
%=&\frac{k/\omega}{1-\frac{k^2\gamma_av_{th,a}^2}{\omega^2}}\left[C_a^2\phi_1+\frac{k\gamma_av_{th,a}^2}{\omega}
%[\frac{1}{2}\frac{k}{\omega}v_{a1}v_{a2}+\frac{1}{2n_{a0}}n_{a2}v_{a1}]\right.\nonumber\\
%&\left.+\frac{1}{2}v_{a1}v_{a2}\right]\nonumber\\
=&\frac{k/\omega}{1-\frac{k^2\gamma_av_{th,a}^2}{\omega^2}}[C_a^2\phi_1+\frac{1}{2}[\frac{k^2\gamma_av_{th,a}^2}{\omega^2}
+1]v_{a1}v_{a2}\nonumber\\
&+\frac{1}{2}\frac{k\gamma_av_{th,a}^2}{\omega}\frac{n_{a2}}{n_{a0}}v_{a1}].
\end{align}
Inserting the fist order result of the velocity $v_{a1}$,
%\begin{align}
%v_{a1}=&\frac{kC_a^2/\omega}{1-\frac{k^2\gamma_av_{th,a}^2}{\omega^2}}[1+
%\frac{k/\omega}{2(1-\frac{k^2\gamma_av_{th,a}^2}{\omega^2})}[\frac{k^2\gamma_av_{th,a}^2}{\omega^2}
%+1]v_{a2}\nonumber\\
%&+\frac{1}{2}\frac{k^2\gamma_av_{th,a}^2/\omega^2}{(1-\frac{k^2\gamma_av_{th,a}^2}{\omega^2})}\frac{n_{a2}}{n_{a0}}]\phi_1\nonumber\\
%=&\frac{kC_a^2/\omega}{1-\frac{k^2\gamma_av_{th,a}^2}{\omega^2}}[1+
%\frac{k^2/\omega^2}{2(1-\frac{k^2\gamma_av_{th,a}^2}{\omega^2})^2}[\frac{k^2\gamma_av_{th,a}^2}{\omega^2}
%+1]\nonumber\\
%&\times[C_a^2\phi_2+\frac{1}{4}(1+\frac{2k^2\gamma_av_{th,a}^2}{\omega^2})v_{a1}^2]\nonumber\\
%&+\frac{1}{2}\frac{k^4\gamma_av_{th,a}^2/\omega^4}{(1-\frac{k^2\gamma_av_{th,a}^2}{\omega^2})^2}[C_a^2\phi_2+\frac{3}{4}v_{a1}^2]]\phi_1\nonumber\\
%=&\frac{kC_a^2/\omega}{1-\frac{k^2\gamma_av_{th,a}^2}{\omega^2}}[1+\frac{k^2C_a^2/\omega^2}{2(1-\frac{k^2\gamma_av_{th,a}^2}{\omega^2})^2}
%[\frac{2k^2\gamma_av_{th,a}^2}{\omega^2}+1]\phi_2\nonumber\\
%&+\frac{k^2/\omega^2}{2(1-\frac{k^2\gamma_av_{th,a}^2}{\omega^2})^2}
%[\frac{1}{4}(\frac{k^2\gamma_av_{th,a}^2}{\omega^2}+1)(1+\frac{2k^2\gamma_av_{th,a}^2}{\omega^2})\nonumber\\
%&+
%\frac{3k^2\gamma_av_{th,a}^2}{4\omega^2}]v_{a1}^2]\phi_1
%\end{align}
%then
one obtains
\begin{align}
%v_{a1}=&\frac{kC_a^2/\omega}{1-\frac{k^2\gamma_av_{th,a}^2}{\omega^2}}[1+\frac{k^2C_a^2/\omega^2}{2(1-\frac{k^2\gamma_av_{th,a}^2}{\omega^2})^2}
%[\frac{2k^2\gamma_av_{th,a}^2}{\omega^2}+1]\phi_2\nonumber\\
%&+\frac{k^2/\omega^2}{8(1-\frac{k^2\gamma_av_{th,a}^2}{\omega^2})^2}
%[1+\frac{6k^2\gamma_av_{th,a}^2}{\omega^2}\nonumber\\
%&+
%\frac{2k^4\gamma_av_{th,a}^4}{\omega^4}]v_{a1}^2]\phi_1\nonumber\\
v_{a1}=&\frac{kC_a^2/\omega}{1-\frac{k^2\gamma_av_{th,a}^2}{\omega^2}}[1+\frac{\frac{k^2C_a^2}{\omega^2}( \frac{2k^2\gamma_av_{th,a}^2}{\omega^2}+1)}{2(1-\frac{k^2\gamma_av_{th,a}^2}{\omega^2})^2}
\phi_2\nonumber\\
&+\frac{\frac{k^4C_a^4}{\omega^4}(1+\frac{6k^2\gamma_av_{th,a}^2}{\omega^2}+
	\frac{2k^4\gamma_av_{th,a}^4}{\omega^4})}{8(1-\frac{k^2\gamma_av_{th,a}^2}{\omega^2})^4}
\phi_{1}^2]\phi_1.
\end{align}
Next, the density perturbation is 
\begin{align}
n_{a1}=&n_{a0}\frac{k^2C_a^2/\omega^2}{1-\frac{k^2\gamma_av_{th,a}^2}{\omega^2}}[
1+\frac{3k^2C_a^2/\omega^2}{2(1-\frac{k^2\gamma_av_{th,a}^2}{\omega^2})^2}\phi_2\nonumber\\
&+
\frac{k^4C_a^4/\omega^2}{8(1-\frac{k^2\gamma_av_{th,a}^2}{\omega^2})^4}
(5+\frac{4k^2\gamma_av_{th,a}^2}{\omega^2})\phi_1^2]\phi_1.\label{eq10}
\end{align}
The Possion equation for $l=1$ is
\begin{equation}
[k^2\lambda_{De}^2+1+\phi_0]\phi_1+\frac{1}{2}\phi_2\phi_1=\sum_{a}Z_a\frac{n_{a1}}{n_{e0}}.
\end{equation}
Substituting the expression of Eq. (\ref{eq10}) into this Possion equation gives
\begin{align}
&[k^2\lambda_{De}^2+1+\phi_0-\sum_{a}\frac{Z_an_{a0}/n_{e0}}{1-\frac{k^2\gamma_av_{th,a}^2}{\omega^2}}\frac{k^2C_a^2}
{\omega^2}]\phi_1\\
=&-\frac{1}{2}\phi_2\phi_1+\sum_a\frac{Z_an_{a0}/n_{e0}}{(1-\frac{k^2\gamma_av_{th,a}^2}{\omega^2})^3}
\frac{3k^4C_a^4}{2\omega^4}\phi_2\phi_1\nonumber\\
&+\sum_a\frac{Z_an_{a0}/n_{e0}(5+\frac{4k^2\gamma_av_{th,a}^2}{\omega^2})}
{(1-\frac{k^2\gamma_av_{th,a}^2}{\omega^2})^5}\frac{k^6C_a^6}{\omega^6}\phi_1^2\phi_1.
\end{align}
Applying the relation $\phi_0\approx-\frac{1}{4}|\phi_1|^2$, and defining
\begin{equation}
C_{6s}^6\equiv\sum_{a}\frac{Z_an_{a0}(1+\frac{4k^2\gamma_av_{th,a}^2}{5\omega^2})}{n_{e0}(1-\frac{k^2\gamma_av_{th,a}^2}{\omega^2})^5}C_a^6,
\end{equation}
one obtains
\begin{align}
&[k^2\lambda_{De}^2+1-\frac{k^2C_{2s}^2}{\omega^2}]\phi_1=
[-\frac{1}{2}+\frac{3k^4C_{4s}^2}{2\omega^4}]\phi_2\phi_1
\nonumber\\
&+[\frac{1}{4}+\frac{5k^6C_{6s}^6}{8\omega^6}]|\phi_1|^2\phi_1.
\end{align}
Substituting $\phi_2=A_{2\phi}\phi_1^2$ and defining $C_{s1}^2=k^2C_{2s}^2/(1+k^2\lambda_{De}^2)$, one can obtain
\begin{align}
[\omega^2-C_{s1}^2]\phi_1=(A_{2\phi}C_{A_{2\phi}}+C_2)|\phi_1|^2\phi_1,
\end{align}
where
\begin{align}
&C_{A_{2\phi}}=\frac{C_{s1}^2}{2}[-\frac{\omega^2}{k^2C_{2s}^2}+
3\frac{k^2C_{4s}^4}{\omega^2C_{2s}^2}],\\
&C_2=\frac{C_{s1}^2}{8}[\frac{2\omega^2}{k^2C_{2s}^2}+\frac{5k^4C_{6s}^6}{\omega^4C_{2s}^2}].
\end{align}
 Since $\omega_{\mathrm{harm}}=\omega+\delta\omega_{\mathrm{harm}}$, and $\omega^2-C_{s1}^2(\omega)=0$, $\omega_{\mathrm{harm}}$ is the effective fundamental IAWs frequency after accounting for harmonic effects.
Due to the inclusion of the second harmonic terms, the frequency shift of the fundamental mode in multi-ion species plasmas is given by
\begin{align}
\frac{\delta \omega_{\mathrm{harm}}}{\omega}=
\frac{\Delta}{2\omega^2-\omega\partial_{\omega}C_{s1}^2}|\phi_1|^2,\quad\Delta\equiv A_{2\phi}C_{A_{2\phi}}+C_2.
\end{align}

\subsection{single-ion species plasmas}
\begin{figure}[tp!]
	\centering
	\includegraphics[width=1\columnwidth]{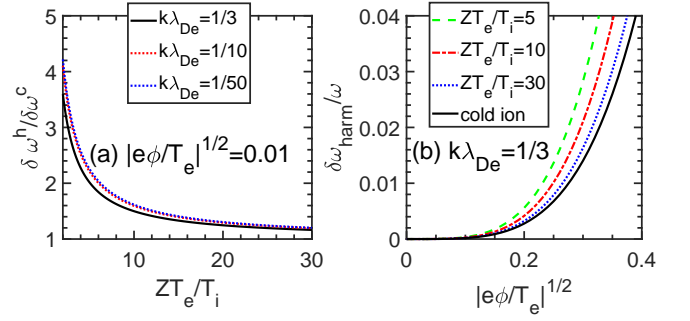}
		\caption{(a) The ratio of the fluid nonlinear frequency shift including ion temperature to that with cold ion varies with $ZT_e/T_i$ for different $k\lambda_{De}$. (b) The fluid nonlinear frequency shift varies with the electrostatic potential for different $ZT_e/T_i$ with $k\lambda_{De}=1/3$.}
		\label{fig1}
\end{figure}

Applying our result for single ion plasmas
\begin{align}
  &\omega^2=k^2C_s^2/\alpha+k^2\gamma_av_{th,a}^2=k^2C_s^2(1/\alpha+\beta),\\
  &A_{2\phi}=\frac{3\alpha^2(1+\alpha\beta)-1}{12(\alpha-1)},\\
  &C_{A_{2\phi}}=\frac{\omega^2}{2}[-\frac{1}{\alpha}+3\alpha(1+\alpha\beta)],\\
  &C_{2}=\frac{\omega^2}{8}[\frac{2}{\alpha}+5\alpha^2(1+\alpha\beta)^2(1+\frac{4k^2\gamma_av_{th,a}^2}{5\omega^2})],\\
  &2\omega^2-\omega\partial_{\omega}C_{s1}^2=2\omega^2(1+\alpha\beta),
\end{align}
where $\alpha=k^2\lambda_{De}^2+1$ and $\beta=\gamma_av_{th,a}^2/C_s^2=\frac{\gamma_aT_a}{ZT_e}$.
Therefore, the fluid nonlinear frequency shift in single ion species plasmas is given by
\begin{align}
  \frac{\delta\omega_{\mathrm{harm}}}{\omega}=&\frac{1}{48\alpha(\alpha-1)(1-\beta)}
  [(3\alpha^2(1+\alpha\beta)-1)^2\nonumber\\
  &+
  3(\alpha-1)[2+5\alpha^3(1+\alpha\beta)^2(1+c)]|\phi_1|^2,\label{eq11}
\end{align}
where $c=\frac{4k^2\gamma_av_{th,a}^2}{5\omega^2}$. For cold ion, $\beta=0,c=0$, the nonlinear frequency shift is
\begin{align}
  \frac{\delta \omega_{\mathrm{harm}}}{\omega}=&
\frac{[(3\alpha^2-1)^2+3(\alpha-1)(2+5\alpha^3)]}{48\alpha(\alpha-1)}|\phi_1|^2\nonumber\\
=&\frac{(4+45\tilde{k}^2+93\tilde{k}^4+81\tilde{k}^6+24\tilde{k}^8)}{48\tilde{k}^2(1+\tilde{k}^2)}|\phi_1|^2,
\label{eq12}
\end{align}
which is consistent with the result of Berger {\it et al.} \cite{Berger_2013POP} from cold ion assumption. Where $\tilde{k}=k\lambda_{De}$ is defined. Comparing Eq. (\ref{eq11}) with Eq. (\ref{eq12}), the effect of ion temperature on the frequency shift is presented by these terms including $\beta$.
It is obvious that ion temperature can enhance the frequency shift. More detail, ion temperature will make a significant effect if $\gamma_aT_a/Z_aT_e\gtrsim0.1$.

As shown in Fig. \ref{fig1}(a), the ratio of the fluid NFS including ion temperature to that with cold ion $\delta\omega^h/\delta\omega^c$ will decrease with $ZT_e/T_i$ increasing. When $ZT_e/T_i$ is very large, such as $ZT_e/T_i=30$, the fluid NFS including ion temperature will be close to the fluid NFS with cold ion. Especially, when $ZT_e/T_i=2$, the ratio $\delta\omega^h/\delta\omega^c$ will reach $3-4$. That is to say, the effect of ion temperature on fluid NFS is obvious in higher ion temperature. On the other hand, with $k\lambda_{De}$ increasing, $\delta\omega^h/\delta\omega^c$ will decrease. When $ZT_e/T_i$ is very large, such as $ZT_e/T_i=30$, $\delta\omega^h/\delta\omega^c$
will be the same in different wave numbers $k\lambda_{De}$, which is because the effect of thermal ions on fluid NFS will not be obvious when $ZT_e/T_i$ is very large.
As shown in Fig. \ref{fig1}(b), the fluid NFS will decrease obviously with $ZT_e/T_i$ increasing. Especially, when $ZT_e/T_i=\infty$ or $T_i/T_e=0$, the NFS from single thermal ion fluid theory will be consistent to NFS from single cold ion fluid theory \cite{Berger_2013POP,Chapman_2013PRL}. However, the higher the ion temperature is, the effect of thermal ions on fluid NFS will be more obvious.

\begin{figure}[htbp]
	\centering
	\includegraphics[width=0.9\columnwidth]{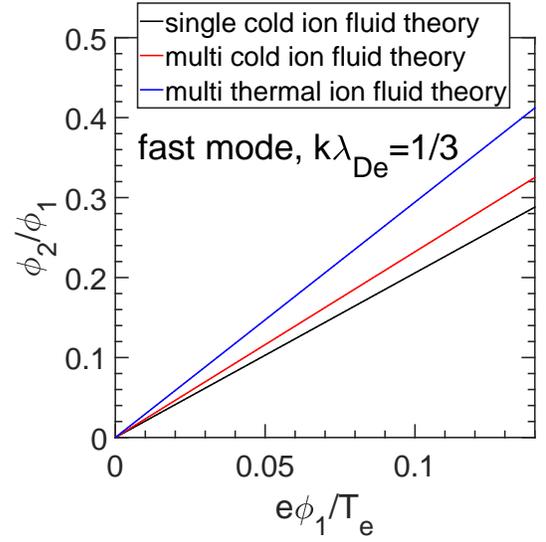}
	\caption{Comparison of the multi thermal ion fluid theory in this paper with the single cold ion fluid theory \cite{Berger_2013POP,Chapman_2013PRL} and multi cold ion fluid theory \cite{Feng_2016PRE}. The condition is $k\lambda_{De}=1/3, T_i/T_e=1/15$ for the fast mode, which is the same as the research of Chapman et al. \cite{Chapman_2013PRL}.}
	\label{fig2}
\end{figure}

\begin{figure}[htbp]
	\centering
	\includegraphics[width=1\columnwidth]{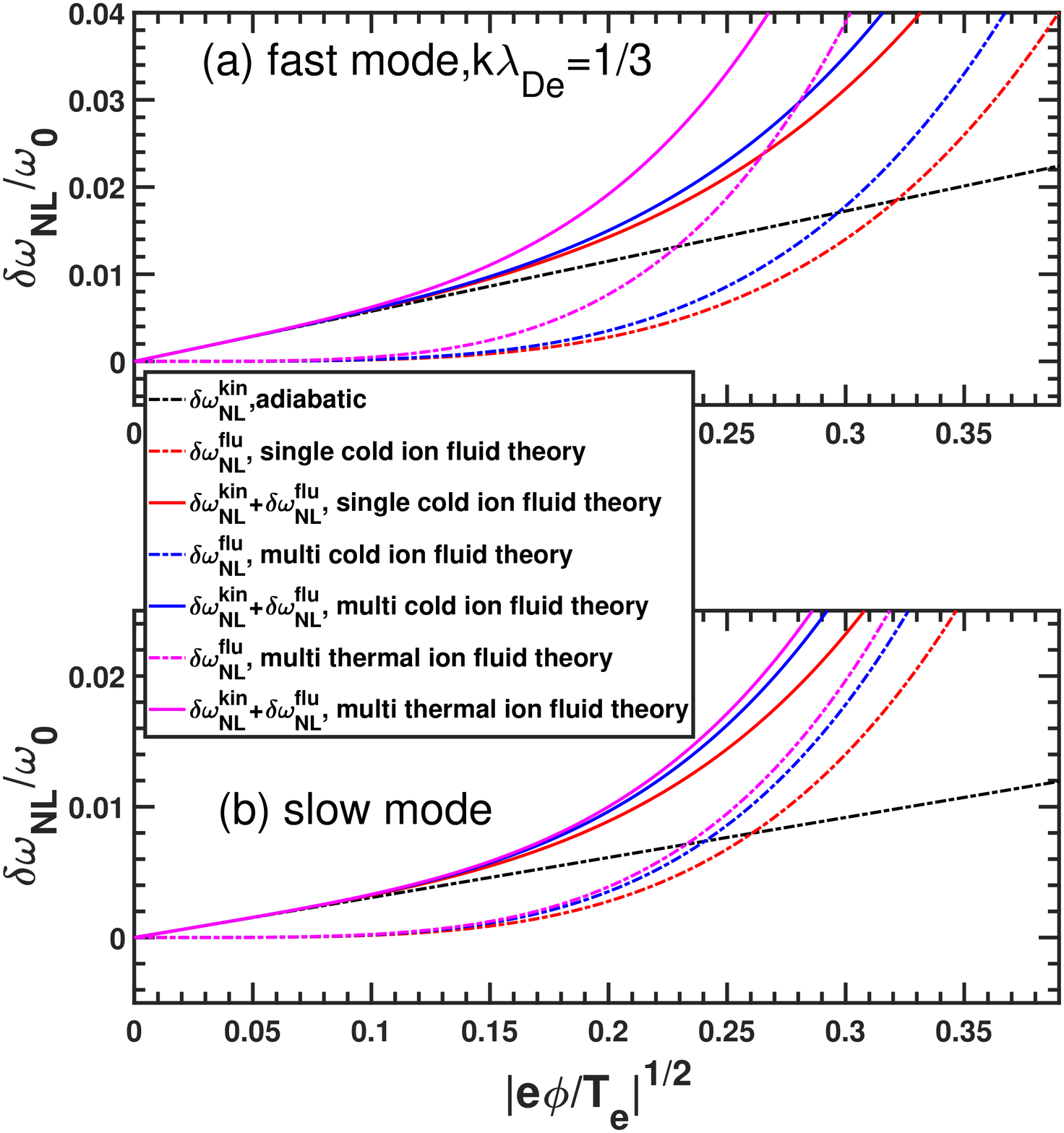}
	\caption{The three theories of fluid nonlinear frequency shift of (a) the fast mode in the condition of $k\lambda_{De}=1/3, T_i/T_e=1/15$ and (b) the slow mode in the condition of $k\lambda_{De}=1/3, T_i/T_e=1/2$. Where the single cold ion fluid theory is shown in Chapman et al.'s research \cite{Chapman_2013PRL,Berger_2013POP}, the multi cold ion fluid theory is given by Feng et al. \cite{Feng_2016PRE}, and the multi thermal ion fluid theory is given in this paper. And the kinetic NFS is given by Berger et al. \cite{Berger_2013POP,Chapman_2013PRL}. }
	\label{fig3}
\end{figure}

\subsection{multi-ion species plasmas}
In multi-ion species plasmas, the linear dispersion relation of IAWs can be calculated  from \cite{Williams_1995POP,Berger_2013POP,Feng_2016POP,Feng_2016PRE}
\begin{equation}
\label{Eq:Dispersion_CH}
\epsilon_L(\omega_s,k_s)=1+\sum_j \frac{1}{(k_s\lambda_{Dj})^2}(1+\xi_jZ(\xi_j))=0,
\end{equation}
where 
$Z(\xi_j)=1/\sqrt{\pi}\int_{-\infty}^{+\infty}e^{-v^2}/(v-\xi_j)dv$ is the plasma dispersion function,
and $\lambda_{Dj}=\sqrt{T_j/4\pi n_jZ_j^2e^2}$, $v_{tj}=\sqrt{T_j/m_j}$ are the Debye length and the thermal velocity of specie $j$. And $m_j, Z_j, T_j, n_j$ are the mass, charge number, temperature and density of specie $j$, respectively. In this paper, CH plasmas will be chosen as a typical example of multi-ion species plasmas due to its potential applications in ICF \cite{He_2016POP,Glenzer_2007Nature,Glenzer_2010Science}. In CH plasmas, calculated by Eq. (\ref{Eq:Dispersion_CH}), the frequency of the fast mode in the condition of $k\lambda_{De}=1/3, T_i/T_e=1/15$ is $Re(\omega_s)=6.102\times 10^{-3}\omega_{pe}$, and the frequency of the slow mode in the condition of $k\lambda_{De}=1/3, T_i/T_e=1/2$ is $Re(\omega_s)=5.664\times 10^{-3}\omega_{pe}$.

Figure \ref{fig2} gives the relation between $\phi_2/\phi_1$ and $\tilde{\phi_1}$ from three theories. It is shown that $A_{2\phi}$ calculated by the multi thermal ion fluid theory in this paper is obviously larger than those calculated by the single cold ion fluid theory \cite{Berger_2013POP,Chapman_2013PRL} and multi cold ion fluid theory \cite{Feng_2016PRE}. This result may give an explanation of that the Vlasov simulation data is much larger than the single cold ion fluid theory in Figure 3(b) in the research of Chapman et al. \cite{Chapman_2013PRL}, when the effect of multi-ion species and thermal ions are considered. The multi thermal ion fluid theory given by this paper will be closer to the Vlasov simulation data than multi cold ion fluid theory \cite{Feng_2016PRE} and single cold ion fluid theory \cite{Berger_2013POP,Chapman_2013PRL}.

Figure \ref{fig3} gives the fluid NFS calculated by three theories. Compared with single cold ion fluid theory and multi cold ion fluid theory, the effect of the thermal ions will give a larger fluid NFS. The Vlasov simulation results of the slow mode in Figure 3(d) in Chapman et al.'s work \cite{Chapman_2013PRL} is obviously larger than the single cold ion fluid theory, especially when $|e\phi/T_e|^{1/2}$ is large, because only the single species and cold ions are considered in Chapman et al.'s research, while the system researched by Chapman et al. is CH plasmas but not single-ion species system and the ion temperature with $T_i/T_e=1/2$ could not be neglected. However, for the fast mode, the ion temperature is very low related to the electron temperature, thus the thermal ion effect can be neglected and the multi cold ion fluid theory can be applied. The multi thermal ion fluid theory in this paper will give a correction to the single cold ion fluid theory \cite{Berger_2013POP,Chapman_2013PRL} and multi cold ion fluid theory \cite{Feng_2016PRE}. The effect of multi-ion species and thermal ions is considered in this paper.

\section{Numerical Results \label{sec3}}

\begin{figure}[tp!]
	\centering
	\includegraphics[width=1.05\columnwidth]{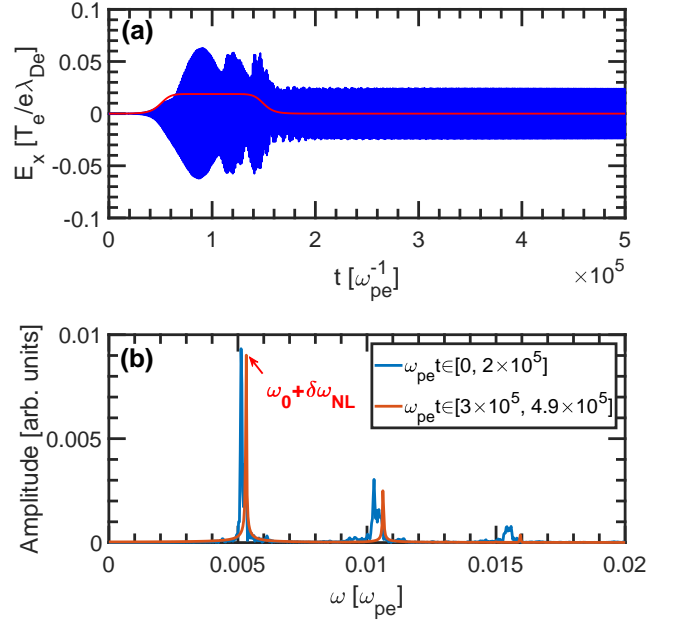}
	\caption{(a) Time evolution of the electric field, calculated at a fixed point $x_0=5\lambda_{De}$ in the condition of $T_i/T_e=1/2$, $k\lambda_{De}=0.3$. Where the red line is the envelop of the driver with the maximum amplitude of $\tilde{E}_d^{max}=0.0188$ and frequency of $\omega_0=5.121\times10^{-3}\omega_{pe}$, which is the linear frequency of the slow mode. (b) Frequency spectra of $E_x$ in different time scopes.
	}
	\label{fig4}
\end{figure}

\begin{figure}[tp!]
	\centering
	\includegraphics[width=1\columnwidth]{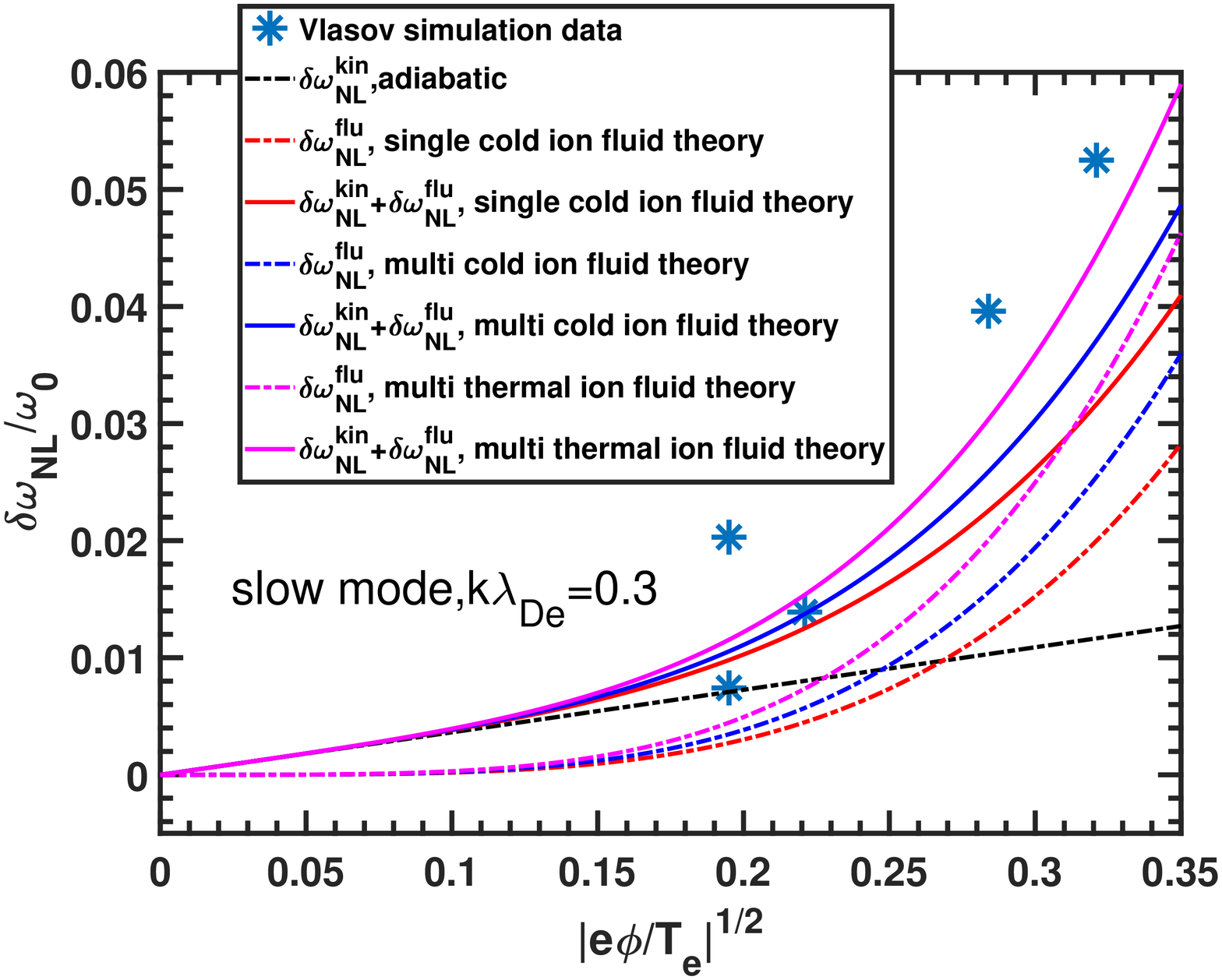}
	\caption{The comparison of Vlasov simulation data with the fluid NFS and kinetic NFS. Where the condition is $k\lambda_{De}=0.3, T_i/T_e=1/2$ for the slow mode. The fluid NFS are given by three different fluid theories and the kinetic NFS is given by Berger et al. \cite{Berger_2013POP,Chapman_2013PRL}.}
	\label{fig5}
\end{figure}

One dimension in space and velocity (1D1V) Vlasov-Poisson code\cite{Liu_2009POP,Liu_2009POP_1} is taken to excite the nonlinear IAW in CH plasmas. The form of the external driving electric field (driver) is
\begin{equation}
\tilde{E}_d(x, t)=\tilde{E}_d(t)\text{sin}(kx-\omega_d t),
\end{equation}
where  $\tilde{E}_d=eE_d\lambda_{De}/T_e$. $\omega_d$ and k are the frequency and the wave number of the driver. The envelope of the driver $\tilde{E}_d(t)$ is
\begin{equation}
\tilde{E}_d(t)=\frac{\tilde{E}_d^{max}}{1+(\frac{t-t_0}{\frac{1}{2}t_0})^{10}},
\end{equation}
where the maximum amplitude of the driver is $\tilde{E}_d^{max}=eE_d^{max}\lambda_{De}/T_e$. And the duration time of the peak driving electric filed is $t_0$. The driver frequency chooses the fundamental frequency of the linear IAW, i.e., $\omega_d=\omega_L$. As shown in Fig. \ref{fig4}(a), the duration time of external driving electric field is $t_0=1\times10^{5}\omega_{pe}^{-1}$ and turns off at $\omega_{pe}t=2\times10^5$.  After the driver is off, the BGK \cite{BGK} mode is established, and the electric field oscillates at almost constant amplitude. As shown in Fig. \ref{fig4}(b), when the driver is on, the frequency of the electric field among the time scope of $\omega_{pe}t\in[0, 2\times10^5]$ is close to the linear frequency of the slow mode $\omega_0$. However, when the driver is off, the electric field will keep constant since the steady BGK \cite{BGK} mode will be established through particle trapping. The nonlinear frequency shift of the slow mode will occur due to particle trapping and harmonic generation. During $\omega_{pe}t\in[3\times10^5, 4.9\times10^5]$, the slow mode with a larger frequency of $\omega_0+\delta\omega_{NL}$ than the linear frequency $\omega_0$ will be established as shown in Fig. \ref{fig4}(b). The nonlinear frequency shift $\delta\omega_{NL}/\omega_0$ is related to the amplitude of nonlinear IAW excited by the driver.

The amplitude of the driver $\tilde{E}_d^{max}$ varies to excite different amplitudes of nonlinear IAW. Therefore, the nonlinear frequency shift $\delta\omega_{NL}/\omega_0$ of the slow mode in different IAW amplitudes under the condition of $k\lambda_{De}=0.3, T_i/T_e=1/2$ can be obtained by Vlasov simulation.
As shown in Fig. \ref{fig5}, the total NFS $\delta\omega_{NL}^{kin}+\delta\omega_{NL}^{flu}$ from multi thermal ion fluid theory is closer to the Vlasov simulation data than that from multi cold ion fluid theory \cite{Feng_2016PRE} and single cold ion fluid theory \cite{Berger_2013POP,Chapman_2013PRL}. That the Vlasov simulation data is also larger than the total NFS from multi thermal ion fluid theory may be because that only the second harmonic terms are considered in the theory in this paper. However, the effect of the thermal ions on fluid NFS would give a better correction to the multi cold ion fluid theory \cite{Feng_2016PRE} and the single cold ion fluid theory \cite{Berger_2013POP,Chapman_2013PRL}, as a result the thermal ion effect should be considered.

\section{\label{sec4}Conclusions}
The effect of thermal ions is considered in the fluid nonlinear frequency shift model and a multi thermal ion fluid model is given. The multi thermal ion fluid model is verified to be better consistent to the Vlasov simulation results and will give a better correction to the previous theory, especially for the slow IAW mode with high ion temperature. It will give a complete theory considering multi-ion species and thermal ions to calculate the frequency of large amplitude nonlinear IAW. This theoretical model will have a potential application in space physics and ICF, since the large amplitude nonlinear IAW always be produced in solar wind and SBS.

\begin{acknowledgments}
We are pleased to acknowledge useful discussions with C. Z. Xiao. This research was supported by National Postdoctoral Program for Innovative Talents (Grant No. BX20180055), the China Postdoctoral Science Foundation (Grant No. 2018M641274), the National Natural Science Foundation of China (Grant Nos. 11875091, 11575035, 11475030 and 11435011) and Science Challenge Project, No. TZ2016005.
\end{acknowledgments}

%\bibliography{vlasovm}

%\bibliographystyle{aipnum4-1}
\bibliographystyle{unsrt}
\bibliography{manuscript1.bib}

\begin{thebibliography}{10}

\bibitem{Vecchio_2014JGR}
A.~Vecchio, F.~Valentini, S.~Donato, V.~Carbone, C.~Briand, J.~Bougeret, and
  P.~Veltri.
\newblock Electrostatic fluctuations in the solar wind: An evidence of the link
  between alfv\'{e}nic and electrostatic scales.
\newblock {\em Journal of Geophysical Research: Space Physics},
  119(9):7012--7024, 2014.

\bibitem{Valentini_2014APJL}
F.~Valentini, A.~Vecchio, S.~Donato, V.~Carbone, C.~Briand, J.~Bougeret, and
  P.~Veltri.
\newblock The nonlinear and nonlocal link between macroscopic alfvénic and
  microscopic electrostatic scales in the solar wind.
\newblock {\em The Astrophysical Journal Letters}, 788(1):L16, 2014.

\bibitem{Gurnett_1977JGR}
Donald~A. Gurnett and Roger~R. Anderson.
\newblock Plasma wave electric fields in the solar wind: Initial results from
  helios 1.
\newblock {\em Journal of Geophysical Research}, 82(4):632--650, 1977.

\bibitem{Gurnett_1978JGR}
D.~A. Gurnett and L.~A. Frank.
\newblock Ion acoustic waves in the solar wind.
\newblock {\em Journal of Geophysical Research}, 83(A1):58--74, 1978.

\bibitem{Gurnett_1979JGR}
D.~A. Gurnett, E.~Marsch, W.~Pilipp, R.~Schwenn, and H.~Rosenbauer.
\newblock Ion acoustic waves and related plasma observations in the solar wind.
\newblock {\em Journal of Geophysical Research}, 84(A5):2029--2038, 1979.

\bibitem{He_2016POP}
X.~T. He, J.~W. Li, Z.~F. Fan, L.~F. Wang, J.~Liu, K.~Lan, J.~F. Wu, and W.~H.
  Ye.
\newblock A hybrid-drive nonisobaric-ignition scheme for inertial confinement
  fusion.
\newblock {\em Physics of Plasmas}, 23(8):082706--, 2016.

\bibitem{Glenzer_2010Science}
S.~H. Glenzer, B.~J. MacGowan, P.~Michel, N.~B. Meezan, L.~J. Suter, S.~N.
  Dixit, J.~L. Kline, G.~A. Kyrala, D.~K. Bradley, D.~A. Callahan, E.~L.
  Dewald, L.~Divol, E.~Dzenitis, M.~J. Edwards, A.~V. Hamza, C.~A. Haynam,
  D.~E. Hinkel, D.~H. Kalantar, J.~D. Kilkenny, O.~L. Landen, J.~D. Lindl,
  S.~LePape, J.~D. Moody, A.~Nikroo, T.~Parham, M.~B. Schneider, R.~P.~J. Town,
  P.~Wegner, K.~Widmann, P.~Whitman, B.~K.~F. Young, B.~Van~Wonterghem, L.~J.
  Atherton, and E.~I. Moses.
\newblock Symmetric inertial confinement fusion implosions at ultra-high laser
  energies.
\newblock {\em Science}, 327(5970):228--1231, 2010.

\bibitem{Glenzer_2007Nature}
S.~H. Glenzer, D.~H. Froula, L.~Divol, M.~Dorr, R.~L. Berger, S.~Dixit, B.~A.
  Hammel, C.~Haynam, J.~A. Hittinger, J.~P. Holder, O.~S. Jones, D.~H.
  Kalantar, O.~L. Landen, A.~B. Langdon, S.~Langer, B.~J. MacGowan, A.~J.
  Mackinnon, N.~Meezan, E.~I. Moses, C.~Niemann, C.~H. Still, L.~J. Suter,
  R.~J. Wallace, E.~A. Williams, and B.~K.~F. Young.
\newblock Experiments and multiscale simulations of laser propagation through
  ignition-scale plasmas.
\newblock {\em Nat. Phys.}, 3(10):716--719, 10 2007.

\bibitem{LanKe_2017PRE}
Ke~Lan, Zhichao Li, Xufei Xie, Yao-Hua Chen, Chunyang Zheng, Chuanlei Zhai,
  Liang Hao, Dong Yang, Wen~Yi Huo, Guoli Ren, Xiaoshi Peng, Tao Xu, Yulong Li,
  Sanwei Li, Zhiwen Yang, Liang Guo, Lifei Hou, Yonggang Liu, Huiyue Wei,
  Xiangming Liu, Weiyi Cha, Xiaohua Jiang, Yu~Mei, Yukun Li, Keli Deng, Zheng
  Yuan, Xiayu Zhan, Haijun Zhang, Baibin Jiang, Wei Zhang, Xuewei Deng, Jie
  Liu, Kai Du, Yongkun Ding, Xiaofeng Wei, Wanguo Zheng, Xiaodong Chen, E.~M.
  Campbell, and Xian-Tu He.
\newblock Experimental demonstration of low laser-plasma instabilities in
  gas-filled spherical hohlraums at laser injection angle designed for ignition
  target.
\newblock {\em Phys. Rev. E}, 95:031202, Mar 2017.

\bibitem{Lan_2016MRE}
Ke~Lan, Jie Liu, Zhichao Li, Xufei Xie, Wenyi Huo, Yaohua Chen, Guoli Ren,
  Chunyang Zheng, Dong Yang, Sanwei Li, Zhiwen Yang, Liang Guo, Shu Li, Mingyu
  Zhang, Xiaoying Han, Chuanlei Zhai, Lifei Hou, Yukun Li, Keli Deng, Zheng
  Yuan, Xiayu Zhan, Feng Wang, Guanghui Yuan, Haijun Zhang, Bobin Jiang, Lizhen
  Huang, Wei Zhang, Kai Du, Runchang Zhao, Ping Li, Wei Wang, Jingqin Su,
  Xuewei Deng, Dongxia Hu, Wei Zhou, Huaiting Jia, Yongkun Ding, Wanguo Zheng,
  and Xiantu He.
\newblock Progress in octahedral spherical hohlraum study.
\newblock {\em Matter and Radiation at Extremes}, 1(1):8 -- 27, 2016.

\bibitem{Huo_2016PRL}
Wen~Yi Huo, Zhichao Li, Yao-Hua Chen, Xuefei Xie, Ke~Lan, Jie Liu, Guoli Ren,
  Yongsheng Li, Yonggang Liu, Xiaohua Jiang, Dong Yang, Sanwei Li, Liang Guo,
  Huan Zhang, Lifei Hou, Huabing Du, Xiaoshi Peng, Tao Xu, Chaoguang Li, Xiayu
  Zhan, Guanghui Yuan, Haijun Zhang, Baibin Jiang, Lizhen Huang, Kai Du,
  Runchang Zhao, Ping Li, Wei Wang, Jingqin Su, Yongkun Ding, Xian-Tu He, and
  Weiyan Zhang.
\newblock First investigation on the radiation field of the spherical hohlraum.
\newblock {\em Phys. Rev. Lett.}, 117:025002, Jul 2016.

\bibitem{Huo_2016MRE}
Wenyi Huo, Zhichao Li, Dong Yang, Ke~Lan, Jie Liu, Guoli Ren, Sanwei Li, Zhiwen
  Yang, Liang Guo, Lifei Hou, Xuefei Xie, Yukun Li, Keli Deng, Zheng Yuan,
  Xiayu Zhan, Guanghui Yuan, Haijun Zhang, Baibin Jiang, Lizhen Huang, Kai Du,
  Runchang Zhao, Ping Li, Wei Wang, Jingqin Su, Yongkun Ding, Xiantu He, and
  Weiyan Zhang.
\newblock First demonstration of improving laser propagation inside the
  spherical hohlraums by using the cylindrical laser entrance hole.
\newblock {\em Matter and Radiation at Extremes}, 1(1):2 -- 7, 2016.

\bibitem{Froula_2002PRL}
D.~H. Froula, L.~Divol, and S.~H. Glenzer.
\newblock Measurements of nonlinear growth of ion-acoustic waves in
  two-ion-species plasmas with thomson scattering.
\newblock {\em Phys. Rev. Lett.}, 88:105003, Feb 2002.

\bibitem{Berger_1998POP}
R.~L. Berger, C.~H. Still, E.~A. Williams, and A.~B. Langdon.
\newblock On the dominant and subdominant behavior of stimulated raman and
  brillouin scattering driven by nonuniform laser beams.
\newblock {\em Physics of Plasmas}, 5(12):4337--4356, 1998.

\bibitem{Neumayer_2008PRL}
P.~Neumayer, R.~L. Berger, L.~Divol, D.~H. Froula, R.~A. London, B.~J.
  MacGowan, N.~B. Meezan, J.~S. Ross, C.~Sorce, L.~J. Suter, and S.~H. Glenzer.
\newblock Suppression of stimulated brillouin scattering by increased landau
  damping in multiple-ion-species hohlraum plasmas.
\newblock {\em Phys. Rev. Lett.}, 100:105001, Mar 2008.

\bibitem{Giacone_1998POP}
R.~E. Giacone and H.~X. Vu.
\newblock Nonlinear kinetic simulations of stimulated brillouin scattering.
\newblock {\em Physics of Plasmas}, 5(5):1455--1460, 1998.

\bibitem{Vu_2001PRL}
H.~X. Vu, D.~F. DuBois, and B.~Bezzerides.
\newblock Transient enhancement and detuning of laser-driven parametric
  instabilities by particle trapping.
\newblock {\em Phys. Rev. Lett.}, 86:4306--4309, May 2001.

\bibitem{Albright_2016POP}
B.~J. Albright, L.~Yin, K.~J. Bowers, and B.~Bergen.
\newblock Multi-dimensional dynamics of stimulated brillouin scattering in a
  laser speckle: Ion acoustic wave bowing, breakup, and laser-seeded
  two-ion-wave decay.
\newblock {\em Physics of Plasmas}, 23(3):032703, 2016.

\bibitem{Bruce_1997POP}
Bruce~I. Cohen, Barbara~F. Lasinski, A.~Bruce Langdon, and Edward~A. Williams.
\newblock Resonantly excited nonlinear ion waves.
\newblock {\em Physics of Plasmas}, 4(4):956--977, 1997.

\bibitem{Rozmus_1992POP}
W.~Rozmus, M.~Casanova, D.~Pesme, A.~Heron, and J.‐C. Adam.
\newblock The local–global analysis of the stimulated brillouin scattering in
  the regime of nonlinear sound waves.
\newblock {\em Physics of Fluids B: Plasma Physics}, 4(3):576--593, 1992.

\bibitem{Cohen_1997POP}
Bruce~I. Cohen, Barbara~F. Lasinski, A.~Bruce Langdon, and Edward~A. Williams.
\newblock Resonantly excited nonlinear ion waves.
\newblock {\em Physics of Plasmas}, 4(4):956--977, 1997.

\bibitem{Berger_2013POP}
R.~L. Berger, S.~Brunner, T.~Chapman, L.~Divol, C.~H. Still, and E.~J. Valeo.
\newblock Electron and ion kinetic effects on non-linearly driven electron
  plasma and ion acoustic waves.
\newblock {\em Physics of Plasmas}, 20(3):032107--, 2013.

\bibitem{Chapman_2013PRL}
T.~Chapman, R.~L. Berger, S.~Brunner, and E.~A. Williams.
\newblock Kinetic theory and vlasov simulation of nonlinear ion-acoustic waves
  in multi-ion species plasmas.
\newblock {\em Phys. Rev. Lett.}, 110:195004, May 2013.

\bibitem{Feng_2016PRE}
Q.~S. Feng, C.~Z. Xiao, Q.~Wang, C.~Y. Zheng, Z.~J. Liu, L.~H. Cao, and X.~T.
  He.
\newblock Fluid nonlinear frequency shift of nonlinear ion acoustic waves in
  multi-ion species plasmas in the small wave number region.
\newblock {\em Phys. Rev. E}, 94:023205, Aug 2016.

\bibitem{Pesme_2005POP}
D.~Pesme, C.~Riconda, and V.~T. Tikhonchuk.
\newblock Parametric instability of a driven ion-acoustic wave.
\newblock {\em Physics of Plasmas}, 12(9):092101, 2005.

\bibitem{Williams_1995POP}
E.~A. Williams, R.~L. Berger, R.~P. Drake, A.~M. Rubenchik, B.~S. Bauer, D.~D.
  Meyerhofer, A.~C. Gaeris, and T.~W. Johnston.
\newblock The frequency and damping of ion acoustic waves in hydrocarbon (ch)
  and two‐ion‐species plasmas.
\newblock {\em Physics of Plasmas}, 2(1):129--138, 1995.

\bibitem{Feng_2016POP}
Q.~S. Feng, C.~Y. Zheng, Z.~J. Liu, C.~Z. Xiao, Q.~Wang, and X.~T. He.
\newblock Excitation of nonlinear ion acoustic waves in ch plasmas.
\newblock {\em Physics of Plasmas}, 23(8):082106--, 2016.

\bibitem{Liu_2009POP}
Z.~J. Liu, S.~P. Zhu, L.~H. Cao, C.~Y. Zheng, X.~T. He, and Yugang Wang.
\newblock Enhancement of backward raman scattering by electron-ion collisions.
\newblock {\em Physics of Plasmas}, 16(11):112703--, 2009.

\bibitem{Liu_2009POP_1}
Z.~J. Liu, X.~T. He, C.~Y. Zheng, and Y.~G. Wang.
\newblock The transition from plasma gratings to cavitons in laser-plasma
  interactions.
\newblock {\em Physics of Plasmas}, 16(9):093108, 2009.

\bibitem{BGK}
Ira~B. Bernstein, John~M. Greene, and Martin~D. Kruskal.
\newblock Exact nonlinear plasma oscillations.
\newblock {\em Phys. Rev.}, 108:546--550, Nov 1957.

\end{thebibliography}

\end{document}